\documentclass{PoS}

\usepackage{latexsym} 	
\usepackage{amsmath}  	
\usepackage{amsbsy}
\usepackage{epsfig}     
\usepackage{cite}

\newcommand{\be}{\begin{equation}}
\newcommand{\ee}{\end{equation}}
\newcommand{\bea}{\begin{eqnarray}}
\newcommand{\eea}{\end{eqnarray}}
\newcommand{\bean}{\begin{eqnarray*}}
\newcommand{\eean}{\end{eqnarray*}}

\newcommand{\Jpsi}{J/\psi}

\title{
\vspace*{-2.7cm}
\begin{flushright}\texttt{\footnotesize
BI-TP 2009/21\\}
\end{flushright}
\vfill
Charmonium correlators and spectral functions at finite temperature }

\ShortTitle{charmonium correlators and spectral functions at finite temperature }

\author{\speaker{H.-T. Ding}, O. Kaczmarek and H. Satz
\\
Fakult\"{a}t f\"{u}r Physik,
 Universit\"{a}t Bielefeld, D-33615 Bielefeld, Germany\\
     E-mail: \email{hengtong.ding,okacz,satz@physik.uni-bielefeld.de}
     }

\author{F. Karsch\\ 
Physics Department, Brookhaven National Laboratory,Upton, NY 11973, USA\\
 Fakult\"{a}t f\"{u}r Physik,
 Universit\"{a}t Bielefeld, D-33615 Bielefeld, Germany\\
        E-mail: \email{karsch@physik.uni-bielefeld.de}}

\author{W. S\"oldner \\
       ExtreMe Matter Institute EMMI, GSI, Planckstr.~1, D-64291 Darmstadt, Germany\\
        E-mail: \email{w.soeldner@gsi.de}}

\abstract{We study charmonium correlators and spectral functions in quenched 
QCD, using Clover improved Wilson fermions on very fine (0.015 fm) 
isotropic lattices at 0.75 $T_{c}$ and $1.5 T_{c}$. We use a new approach to 
distinguish the zero mode contribution from the other contributions. 
Once this is removed, we find that the ratios of correlators to 
reconstructed correlators remain almost unity at all distances. 
The ground state peaks of spectral functions obtained at 0.75 $T_{c}$ are 
reliable and robust. The present accuracy and limited number of points in the temporal direction
at 1.5 $T_c$ do not allow for a reliable conclusion about a possible melting of charmonium states
in the QGP.
}

\FullConference{The XXVII International Symposium on Lattice Field Theory\\
		 July 26-31, 2009\\
		 Peking University, Beijing, China}
\usepackage{epsfig}
\usepackage{graphicx}
\usepackage{dcolumn}
\usepackage{bm}

\begin{document}

\maketitle


One of the longstanding main objectives of heavy-ion physics is to explore the properties of Quark Gluon Plasma (QGP), in which the hadronic degrees of freedom are deconfined. The dissociation of charmonium is proposed to be a signal of the formation of QGP due to the electro-color screening~\cite{Satz:86}. Experimentally, both at RHIC and SPS the suppression of $\Jpsi$ has been observed~\cite{PHENIX:08}, however its interpretation is still not quite understood. 
 To fully understand this phenomenon, a detailed knowledge on the behavior of charmonium states and their dissociation temperatures is of fundamental importance for present and upcoming heavy ion collisions.  From the theoretical point of view, the mesonic spectral function at finite temperature, which has all the information of the hadron properties in the thermal medium, e.g.  transport properties as well as dissociation temperatures, is the key quantity to be investigated.  
 
  Due to its success at the zero temperature, the potential model is applied to this phenomenon~\cite{Reviews:09}, based either on models or finite temperature lattice QCD results for the heavy quark potential in a non-relativistic Schr\"odinger equation. The output dissociation temperature depends strongly on the potential used. Thus it it is important to have a first-principle calculation of dissociation temperature of charmonium in the hot medium. With the lattice QCD approach, the properties of the charmonium, which can be directly seen from the spectral function, are contained in the Euclidean time correlation functions. The extraction of spectral functions from correlators is rather difficult due to the limited number of points in temporal direction required to perform an analytic continuation from imaginary to real time. The most common used method to obtain spectral function is the Maximum Entropy Method (MEM)~\cite{Asakawa:01}. It is based on the Bayesian algorithm and requires the prior knowledge of spectral functions as an input. Our first aim is to discuss the algorithm and emphasize the reasonable prior knowledge to be used. Recently, the zero mode contribution was argued to be the main contribution to the temperature dependence of the spectral function and  one should get rid of it in MEM analysis on the deformation of the spectral function~\cite{Umeda:07}. We propose an approach to disentangle this constant zero mode contribution from the others in MEM analysis. By using very fine isotropic lattice with clover improved Wilson action, we study the temperature dependence of $\Jpsi$ and discuss the reliability of the output spectral function at 1.5 $T_{c}$.

 The Matsubara correlator calculated on lattice is:
 \be
G_{H}(\tau,T)=\sum_{\vec{x}} \langle~ J_H(\tau,\vec{x})~J_{H}^{\dag}(0,\vec{0})~\rangle_T .
\ee
 $J_H$ is a suitable mesonic operator, here we consider local operator of $\bar{q}(\tau,\vec{x})\Gamma_{H} q(\tau,\vec{x})$, where $\Gamma_{H}=1,\gamma_{\mu},\gamma_{5},\gamma_{5}\gamma{_\mu}$ for scalar, vector, pseudo-scalar and axial vector, respectively. The temperature T is related to Euclidean temporal extent a$N_{\tau}$ by T=1/(aN$_{\tau}$), where a is the lattice spacing. Through analytic calculation, the Matsubara correlator can be related to the hadronic spectral function as the following:
\be
G_{H}(\tau,T)=\int_0^{\infty}{\mathrm{d}\omega~\sigma_{H}(\omega,T)}~K(\tau,T,\omega),
\label{eq:relation_cor_spf}
\ee
where the kernel K is given by
\be
K(\tau,T,\omega)= \frac{\mathrm{cosh}(\omega(\tau-\frac{1}{2T}))}{\mathrm{sinh}(\frac{\omega}{2T})}.
\ee

Inverting Eq.~\ref{eq:relation_cor_spf} to extract the spectral function at finite temperature lattice QCD is hampered mainly by two issues: the temporal extent is always restricted by the temperature, $a\tau\leq1/T$;  the spectral functions we want to have should be continuous and have a degree of freedom of $\mathcal{O}$(1000) but the correlators are calculated in the discretized time slices with limited numbers, typically $\mathcal{O}$(10), which makes the inversion  ill-posed.

The basic idea of common used MEM algorithm is to get the most probable spectral function in given data by maximizing the conditional probability 
$Q(\sigma;\alpha)=\mathrm{exp}(\alpha S[\sigma] - L[\sigma])$,
 where $L[\sigma]$ is the usual likelihood function and minimized in the standard $\chi^2$ fit, and the Shannon-Jaynes entropy $S[\sigma]$ is defined as
\be
S[\sigma] = \int_0^{\infty}{\mathrm{d}\omega~\left[\sigma(\omega) - m(\omega) - \sigma(\omega)\mathrm{log}\left(\frac{\sigma(\omega)}{m(\omega)}\right)\right]},
\ee
where $m(\omega)$ is the default model and it requires the prior information of spectral function $\sigma(\omega)$ as input. $\alpha$ is a real and positive parameter which controls the relative weight of the entropy S and the likelihood function L.

There are two important remarks about MEM~\cite{Ding:09}:
\begin{itemize}
 \item No matter what default model one is using, the correlators calculated from the spectral functions obtained from MEM analysis always reproduce the lattice correlator data within the errors,
  \item The spectral functions obtained from MEM always reproduce the high energy behavior of the default model.
\end{itemize}
The first remark mainly concerns the quality of data, which requires huge computing time to get high statistics and large number of data points; the second remark requires correct high energy information of the spectral function to be provided in the default model. 

As the default model is a very important parameter in MEM analysis, one should put reasonable information into it. The high energy behavior of the spectral functions, due to asymptotic freedom, should resemble the free case. Most of the present studies are employing the free continuum spectral as the prior knowledge~\cite{Datta:04,Asakawa:04,Umeda:05,Jakovac:07,Aarts:07spf}, which has the following form~\cite{freespf}:
\bea
\sigma_{H}  &=& \frac{N_{c}}{8\pi^2}\Theta(\omega^2-4m^2)~\omega^2\mathrm{tanh}(\frac{\omega}{4T}) 
				 \sqrt{1-\left(\frac{2m}{\omega}\right)^2} \nonumber \\
				 & &\times \Big{[} a_{H} + \left(\frac{2m}{\omega}\right)^2~ b_H\Big{]} + ~\frac{N_{c}}{3}\frac{T^2}{2}~f_H~\omega\delta(\omega),
				 \label{free_cont_spf}
 \eea
 where $N_{c}$ is number of colors, m is the mass of quark, $a_H$, $b_H$ and $f_H$ are the specified coefficients for mesonic channels. On the lattice, the free spectral function is distorted due to the discretization effects, which mainly show up in the high energy region and has characteristic cusp structures and automatic smooth cutoff at $\omega/T\simeq\mathrm{log}7$. Thus it is absolutely necessary for MEM to include free lattice spectral functions instead of free continuum spectral functions into default model. 
 
    The very low energy part of the spectral functions, due to its relation with transport properties of QGP, is of intrinsic interest. According to the Kubo formulae, transport coefficients such as electrical conductivity are proportional to the slope of the vector spectral function at vanishing energy. In the free continuum case, as one can see from Eq. \ref{free_cont_spf}, there is a $\omega\delta(\omega)$ term in specified channels, which corresponds to a constant contribution to the correlator.  In the interacting case, the delta function is smeared into a transport peak~\cite{Teaney:06}:
   $\frac{\omega\gamma}{\omega^2 + \gamma^2}$ ,
where $\gamma$ is the width of the peak. This is really another prior knowledge one needs to put into default model in MEM analysis at finite temperature.

To distinguish the transport contribution from the other parts, instead of doing MEM analysis on the correlators themselves, we propose to look into the differences of the neighboring correlators, $G(\tau) - G(\tau+1)$, which gives
\be
G(\tau) -  G(\tau+1) = \int_0^{\infty} \mathrm{d}\omega~\sigma(\omega,T)~ \tilde{K}(\tau,\omega),
\label{eq:subtraction}
\ee
where $\tilde{K}$ is given by
\be   
\tilde{K}(\tau,\omega) =2~\mathrm{sinh}\left(\frac{\omega}{2}\right)\frac{~\mathrm{sinh}(\omega(\frac{N_{\tau}}{2}-\tau-\frac{1}{2}))}{\mathrm{sinh}(\frac{\omega N_{\tau}}{2})}.
\ee
Since spectral function is $\tau$ independent, the relation of Eq.~\ref{eq:subtraction} is exact and without any approximation. Consequently, the spectral function obtained from the inversion of Eq.~\ref{eq:subtraction} should be same as that in Eq.~\ref{eq:relation_cor_spf} except the constant contribution is removed. The new kernel $\tilde{K}$ goes smoothly to zero when energy goes to zero as
\be
\lim_{\omega\rightarrow 0} \tilde{K}(\omega,\tau) = \frac{N_{\tau}- 2\tau-1}{N_{\tau}}\omega~ + ~\mathcal{O}(\omega^3).
\ee
It avoids the divergence problem of the standard kernel at $\omega = 0$ as pointed out by Aarts et al.~\cite{Aarts:07} and can explore the information of spectral functions in very low energy region. 
 
 We now discuss the MEM analysis of correlators mainly in the vector channel. The lattice correlator data are obtained using non-perturbatively improved clover Wilson fermions on isotropic quenched configurations. The lattice parameters are shown in Table ~\ref{tab:lattice_parameters}. We have one coarse lattice (a=0.031fm) and one fine lattice (a=0.015 fm), both of which have temperatures below and above $T_{c}$. The mass of vector meson is tuned to the physical $\Jpsi$ mass. On the fine lattice $am_{c} \approx 0.0987\ll 1$ makes discretization effects small. During  the MEM analysis, we implicitly look at the correlators themselves. We use number of points in the investigated energy region $N_{\omega}=8000$, remove the first two data points and implement the modified kernel $K^{\ast}=\mathrm{tanh}(\omega/2)\cdot K$ to explore the low energy behavior of spectral function.

  	   \begin{table}[ht]
              \centering
              \caption{Lattice parameters}
            \begin{tabular}{c c c  c  c c }
              \hline\hline
	      $\beta$ & $\kappa$ &  a$^{-1}$~[GeV]  & $N_{\sigma}^3 \times N_{\tau}$ & $T/T_c$ &  No. of conf.  \\
	      6.872   & 0.13035  & 6.432    & $128^3 \times 32$              & 0.75    &  126         \\
	              &          &          & $128^3 \times 16$              & 1.5     &  198         \\
              7.457   & 0.13179  & 12.864    & $128^3 \times 64$              & 0.75    &  179         \\
	              &          &          & $128^3 \times 32$              & 1.5     &  108          \\ 
              \hline
            \end{tabular}
            \label{tab:lattice_parameters}
	    \end{table}

\begin{figure}[t]
 \centerline{\epsfig{figure=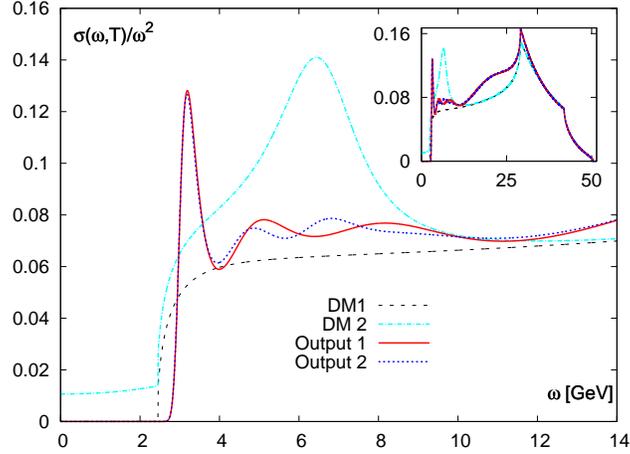,height=6.3cm}}
 \caption{ Default model dependences of the vector spectral function at 0.75 $T_c$ on the fine lattice. The small plot inside is the behavior for the whole energy region.
  }
 \label{fig:DM_dependence}
 \end{figure}

 \begin{figure}[t]
 \centerline{\epsfig{figure=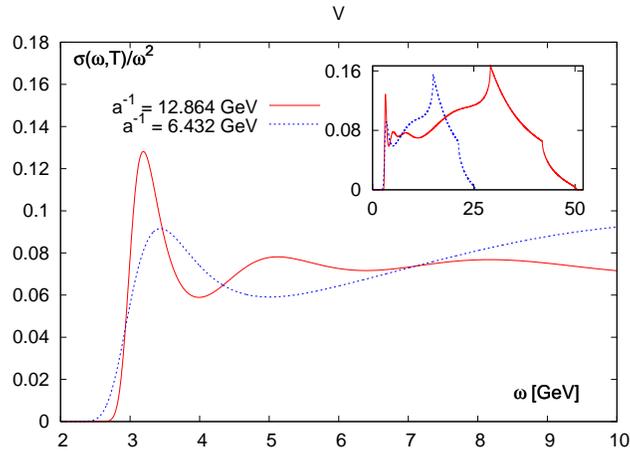,height=6.3cm}}
 \caption{ 
 Comparison of vector spectral functions obtained from coarse and fine lattice at 0.75 $T_{c}$. The 
 small plot inside is the behavior for the whole energy region.
 }
 \label{fig:beta_comparison}
\end{figure}

 In Fig.~\ref{fig:DM_dependence} we show the default model dependence of vector spectral functions at 0.75 $T_{c}$ on the fine lattice. We tried default models with free lattice spectral function (DM 1) and free spectral function plus a resonance peak in the intermediate energy region (DM 2). As one can see, the outputs of MEM reproduce the high energy behavior of the DMs. Although there are some minor differences of the second and third peak from different inputs, the ground state peak is always there and stay clear and robust.  The second and third peak in the  Fig.~\ref{fig:DM_dependence}  could be a mixture of higher states or MEM artifacts due to the finite lattice spacing and limited number of correlator points. The comparison of vector spectral functions obtained from coarse and fine lattice at 0.75 $T_{c}$ in Fig.~\ref{fig:beta_comparison} shows that, when the lattice spacing is reduced, the low energy part, which is of most interest, can be well separated from the high energy part, which contains most of the lattice artifacts. The improvement of output spectral functions can be truly seen with smaller lattice spacing and one could get a hope to get the higher states with the smaller lattice spacing. No zero mode contribution was observed at this temperature.

 \begin{figure}[t]
 \centerline{\epsfig{figure=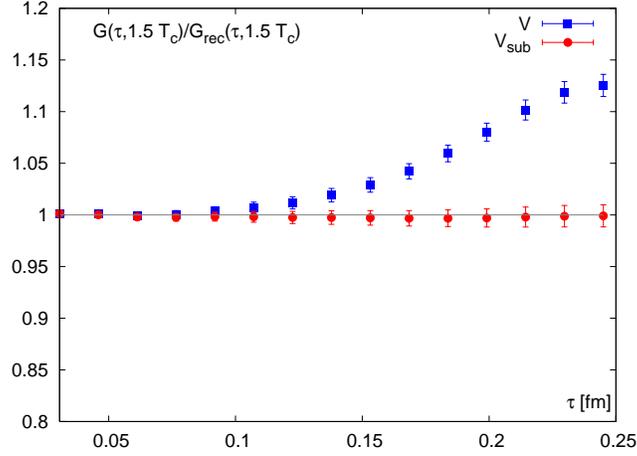,height=6.3cm}}
 \caption{ 
 G/G$_{rec}$ in the vector channel at 1.5 $T_{c}$ to the reconstructed one from 0.75 $T_{c}$ with $\beta$=7.457.
 }
 \label{fig:rec}
\end{figure}

 \begin{figure}[t]
 \centerline{\epsfig{figure=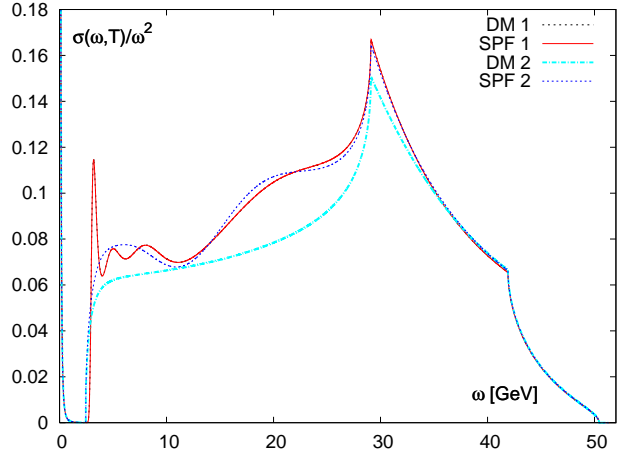,height=6.3cm}}
 \caption{ 
 Default model dependences of vector spectral function at 1.5 $T_{c}$ on the fine lattice.
 }
 \label{fig:spf_1p5Tc}
\end{figure}

   Before we move on to MEM analysis on the temperature above $T_{c}$, which is more complicated due to smaller number of points and smaller physical extent, we investigate the temperature dependence of the vector channel by looking at the ratio of correlators at 1.5 $T_c$ to reconstructed correlators,
 \be
 G_{rec}(\tau,1.5T_c) = \int_{0}^{\infty} \mathrm{d}\omega ~K(\tau,1.5 T_c,\omega)~\sigma(\omega,0.75 T_{c}).
  \ee
One can see from Fig.~\ref{fig:rec} the ratio remains unity at small distances, which is mainly contributed from the high energy part of spectral function, and deviates from unity up to $\approx$12\% at the largest distance, which is dominated by the low energy behavior of the spectral function and indicates the changes in this region. 
 
   For MEM analysis at 1.5 $T_{c}$, we focus on the fine lattice. In Fig.~\ref{fig:spf_1p5Tc}, we tried default models with spectral function obtained from MEM at 0.75 $T_{c}$ plus transport peak (DM 1) and free lattice spectral function plus transport peak (DM 2). SPF 1 and 2 are the corresponding outputs from MEM, respectively. The default model dependence is huge: from SPF 1, we can say $\Jpsi$ is still there with negligible modifications; from SPF 2, probably, $\Jpsi$ melts at 1.5 $T_{c}$. Thus, based on the current data with number points of 32 in the temporal direction, it's hard to tell whether $\Jpsi$ is still there or already dissolved. Furthermore, we calculate the contribution of the transport peak to the correlators, once this is removed, the ratio of correlator data to the reconstructed correlator remains unity within the errors as one can see it from Fig.~\ref{fig:rec} (similar results were obtained in Ref.~\cite{Petreczky:08} through the ratio of time derivative of correlator to time derivative of G$_{rec}$), which indicates the temperature dependence of the correlators could be dominated by the low frequency part of the spectral functions. Then, by using the spectral function obtained from 0.75 $T_{c}$ as default model, we looked into the difference of the neighboring correlator (Eq.~\ref{eq:subtraction}). The output turns out to be exactly the same as the input spectral function at 0.75 $T_{c}$, same as SPF 1 in Fig.~\ref{fig:spf_1p5Tc} with very low energy part suppressed. This accents that the deviation in Fig.~\ref{fig:rec} could be mainly caused by the zero mode contribution.

In summary, we remark the main properties of MEM and stress to put reasonable prior knowledge into 
default model. At 0.75 $T_{c}$, the ground peak of vector spectral function are stable and robust. It is possible to resolve the higher states if the lattice spacing is reduced more. At 1.5 $T_{c}$, It is still hard to tell the fate of $\Jpsi$ in the medium with $N_{\tau} = 32$ and the present statistics. The temperature dependence of spectral function seen from the ratio of correlator data to the reconstructed correlator can be interpreted to be mainly from the zero mode contribution, which can be distinguished from the new approach of MEM analysis.


\begin{acknowledgments}

This work has been supported by the Deutsche Forschungsgemeinschaft under grant GRK 881. The work of WS was supported by the Alliance Program of the Helmholtz Association (HA216/EMMI).

\end{acknowledgments}


\end{document}